\documentclass[journal,twoside,web]{ieeecolor}

\usepackage{array}
\usepackage{tmi}
\usepackage{times}
\usepackage{epsfig}
\usepackage{graphicx}
\usepackage{amsmath}
\usepackage{amssymb}
\usepackage{mathrsfs}
\usepackage{algorithm}
\usepackage{float}
\usepackage{booktabs}
\usepackage{listings}
\usepackage{subfiles}
\usepackage{array}
\usepackage{multirow}
\usepackage{algorithmic}
\usepackage{soul}

\usepackage{cite}

\usepackage[pagebackref=False,breaklinks=True,colorlinks,bookmarks=True]{hyperref}

\def\eg{\emph{e.g.,} }

\newcommand{\bheading}[1]{{\noindent{\textbf{#1}}}}

\def\BibTeX{{\rm B\kern-.05em{\sc i\kern-.025em b}\kern-.08em
    T\kern-.1667em\lower.7ex\hbox{E}\kern-.125emX}}
\begin{document}

\title{Adversarial Medical Image with Hierarchical Feature Hiding}

\author{Qingsong Yao, Zecheng He, Yuexiang Li, Yi Lin, Kai Ma, Yefeng Zheng,~\IEEEmembership{Fellow,~IEEE} \\ and
S. Kevin Zhou,~\IEEEmembership{Fellow,~IEEE}
\thanks{Zhou  is  with  School  of  Biomedical  Engineering \& Suzhou Institute for Advanced Research,  University  of Science  and  Technology  of  China  and  Institute  of  Computing  Technology, Chinese  Academy  of  Sciences. Zhou is the corresponding author. Email: skevinzhou@ustc.edu.cn}
\thanks{Yao is with Institute of Computing Technology, Chinese Academy of Science. Email: yaoqingsong19@mails.ucas.edu.cn}
\thanks{Li, Ma and Zheng are with Jarvis Research Center, Tencent YouTu Lab, Shenzhen, P.R. China. Emails: {\{vicyxli, kylekma, yefengzheng\}}@tencent.com}
\thanks{Li is also with Medical AI ReSearch (MARS) Group, Guangxi Key Laboratory for Genomic and Personalized Medicine, Guangxi Medical University, Nanning, P.R. China.}
\thanks{Lin is with the Department of Computer Science and Engineering, The Hong Kong University of Science and Technology.}
\thanks{He is with Meta Reality Labs. Email:  zechengh@princeton.edu}
}

\maketitle

\begin{abstract}
Deep learning based methods for medical images can be easily compromised by adversarial examples (AEs), posing a great security flaw in clinical decision-making. It has been discovered that conventional adversarial attacks like PGD which optimize the classification logits, are easy to distinguish in the feature space, resulting in accurate reactive defenses. To better understand this phenomenon and \ul{reassess the reliability of the reactive defenses for medical AEs}, we thoroughly investigate the characteristic of conventional medical AEs. Specifically, we first theoretically prove that conventional adversarial attacks change the outputs by continuously optimizing vulnerable features in a fixed direction, thereby leading to outlier representations in the feature space. Then, a stress test is conducted to reveal the vulnerability of medical images, by comparing with natural images. Interestingly, this vulnerability is a double-edged sword, which can be exploited to hide AEs.
We then propose a simple-yet-effective hierarchical feature constraint (HFC), a novel add-on to conventional white-box attacks, which assists to hide the adversarial feature in the target feature distribution. The proposed method is evaluated on three medical datasets, both 2D and 3D, with different modalities. The experimental results demonstrate the superiority of HFC, \emph{i.e.,} it bypasses an array of state-of-the-art adversarial medical AE detectors more efficiently than competing adaptive attacks\footnote{Our code is available at \url{https://github.com/qsyao/Hierarchical_Feature_Constraint}.}, which reveals the deficiencies of medical reactive defense and allows to develop more robust defenses in future.
\end{abstract}


\section{Introduction}

\IEEEPARstart{D}{eep} neural networks (DNNs) are vulnerable to adversarial examples (AEs)~\cite{szegedy2013intriguing}, which are maliciously computed by adding human-imperceptible perturbations to clean images. Those AEs can compromise a network to generate the attacker-desired incorrect predictions~\cite{dong2018boosting}. Accordingly, the adversarial attack in medical imaging analysis~\cite{zhou2021review} is disastrous. For example, the attacker can manipulate the diagnosis of patients in order to raise malicious competition between AI software vendors, threaten the patient's health~\cite{mangaokar2020jekyll}, and even cause public distrust of the DNN-based diagnosis system. More disturbingly, it has been investigated that medical DNNs~\cite{zhou2015medical,zhou2017deep,zhou2019handbook}, including organ segmentation~\cite{ozbulak2019impact}, disease diagnosis~\cite{paschali2018generalizability,finlayson2018adversarial,ma2020understanding}, and landmark detection~\cite{yao2020miss}, are more vulnerable to adversarial attacks than those for natural images.

\begin{figure}[t]
\begin{center}
   \includegraphics[width=1\linewidth]{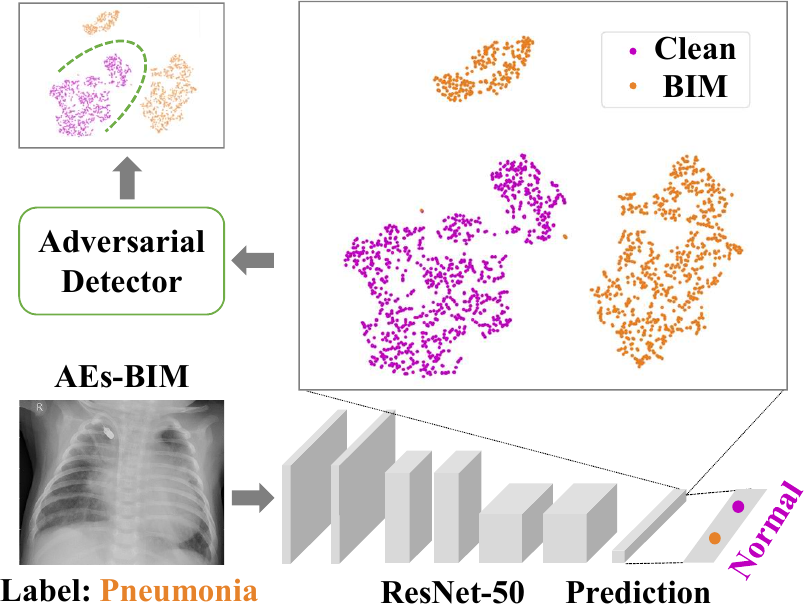}
\end{center}
   \caption{We craft adversarial examples by the state-of-the-art basic iterative method (BIM) \cite{bim} under a small constraint $L_\infty=1$ to manipulate the medical diagnosis result. Then, we visualize the penultimate layer's deep representations of the clean and adversarial examples by 2D t-SNE. Obviously, the adversarial feature lies in an outlier position that can be easily detected by adversarial detectors.}
\vspace{-0.6cm}
\label{Fig:Main}
\end{figure}

Nevertheless, conventional medical AEs like PGD~\cite{PGD} which optimize the classification logits can be easily detected in the feature space~\cite{li2020robust}. In Fig.~\ref{Fig:Main}, we plot a 2D t-SNE~\cite{t-SNE} to illustrate the differences between adversarial and clean features from the penultimate layer of a well-trained pneumonia classifier~\cite{resnet}. It can be observed that adversarial attacks move the feature from the original clean distribution to extreme outlier positions. Therefore, reactive defenders~\cite{ma2018characterizing} can easily make use of this characteristic for the identification of AEs~\cite{MAHA}. 

Since the development process of adversarial attack and defense approaches is an alternate progress, stronger attacks inspire more robust defenses and vice versa, which will both lead to the final robust solution in the future. In this paper, \textbf{we investigate the underlying reason for this phenomenon and reassess the robustness of the reactive defenses}. We raise two important questions. The first one is:
\textit{Why are medical AEs easier to be detected, compared to AEs of natural images?} To further reveal the underlying mechanism, we conduct both theoretical and empirical analyses.  
First, we theoretically prove that the adversarial features are optimized in a consistent direction towards outlier points, where the clean features rarely reside. This applies to both natural and medical images. Then, we conduct a stress test, which aims to change the features by adversarial attacks, and find that medical features are more vulnerable than natural ones in this test. \textbf{Hence, the adversarial features of medical images are pushed more drastically to a consistent direction, and become outliers to a more severe extent than those of natural images, which are easier to distinguish.}

The second question is: \textit{Is it possible to hide a medical AE from being detected by existing defenders in the feature space?} A representative adaptive attack chooses a guiding example and pushes the feature of AE to be close to that of the guiding example~\cite{feature_iclr}. However, there is a limitation when directly applying representative adaptive attacks to medical AEs: different medical images contain different backgrounds and lesion areas, which makes it difficult to manipulate the AE features to be the same to the guiding one in all layers within the perturbation budget. To hide the adversarial feature with a low detectability, we propose a simple-yet-effective hierarchical feature constraint (HFC) as a novel add-on term, which can be plugged into any existing \textit{white-box attack}.
Our HFC first models the clean feature distributions with a Gaussian mixture model for each layer, then encourages adversarial features to move towards the high-density area by maximizing the log-likelihood.

Extensive experiments are conducted on different datasets, including two public datasets on 2D images and one private dataset on 3D volumes, with different backbones and perturbation budgets to validate the effectiveness of our HFC. The experimental results show that the proposed HFC helps medical AEs compromise the state-of-the-art (SOTA) adversarial detectors, and keeps the perturbations small and unrecognizable to humans, which reveals the deficiencies of these medical reactive defenses that only detect medical AEs using OOD metrics. Moreover, we demonstrate that HFC greatly outperforms other adaptive attacks~\cite{feature_iclr,CW_bpda,CW_ten} on manipulating adversarial features. Our experimental results support that \textbf{the greater vulnerability of medical features allows the attacker more room for malicious manipulation}. we believe that our work inspires future defenses that focus on improving the robustness of vulnerable medical features to become more robust.

Overall, we highlight the following contributions:
\begin{itemize}
    \item We theoretically and experimentally investigate the underlying reason to explain why medical AEs are more easily detected in feature space, compared with natural AEs.
    \item We propose the hierarchical feature constraint (HFC), a novel plug-in that can be applied to existing white-box attacks to avoid being spotted.
    \item Extensive experimental results validate the effectiveness of our HFC to help conventional adversarial attacks bypass an array of state-of-the-art adversarial detectors within a small perturbation budget, which exposes the limitations of reactive medical defense and offers insights into future stronger defenses.
\end{itemize}

A preliminary version~\cite{yao2020hierarchical} of this work was previously published as a conference paper.\footnote{The reviewers' comments can be found in \url{https://miccai2021.org/openaccess/paperlinks/2021/09/01/017-Paper1101.html}.} This paper substantially extends the preliminary version with the following improvements: 

\begin{itemize}
    \item We provide a detailed proof of the theorem about feature vulnerability for the binary class setting. Furthermore, we expand the theorem to multi-class and provide both empirical and theoretical analyses for the out-of-distribution adversarial feature in this new multi-class setting.
    \item We additionally evaluate the performance of our HFC on a multi-class disease classification dataset consisting of 3D medical volumes, as well as with different baseline models. In addition, we find that 3D medical volumes are more vulnerable than 2D medical images.
    \item We provide more detailed discussion of experimental results, hyper-parameter analysis, and compare the performances of HFC on medical and natural image datasets, and reveal the importance of the robustness of medical deep representations.
    \item We further evaluate the potential of our HFC term to detect the real outlier signals, by utilizing the HFC term to serve as the metric of the anomaly score, which can detect the abnormal features caused by the synthesized anomaly signals.
\end{itemize}

\section{Related Work}

Given a clean input image $x$ with its ground truth label $y \in [1,2,\ldots,K]$, a DNN classifier $h$ with pretrained parameters $\theta$ 
predicts the class label $y'$ of the input $x$ via:
\begin{align}
    y' &= \mathop{\arg\max}\limits_k~p_\theta(k|x) \equiv \frac{\mathop{\exp}(l_k(x,\theta))} {\sum_{j=1}^K\mathop{\exp}(l_{j}(x,\theta))},
\label{Eq:inference}
\end{align}
where the logit output with respect to class $k$, $l_k(x,\theta)$, is given as $l_k(x,\theta)= \sum_{n=1}^N{w_{nk}*z_n(x, \theta)} + b_k$, in which $z_n(x, \theta)$ is the $n^{th}$ activation of the penultimate layer with $N$ dimensions; $b_k$ and $w_{nk}$ are the bias and the weights from the final dense layer, respectively; and
the $p_\theta(k|x)$ is the probability of $x$ belonging to class $k$. 

A common way of crafting an AE is to manipulate the classifier's outputs by minimizing\footnote{In this work, we focus on targeted adversarial attack that aims to manipulate the output class for an input to a desired label, rather than just an incorrect label.} the classification error between the outputs and the target class $c$, while keeping the AE $x_{adv}$ within a small $\epsilon$-ball of the $L_p$-norm~\cite{PGD} centered at the original clean input $x$, i.e., $\|x_{adv} - x\|_p \leq \epsilon$, where 
$\epsilon$ is perturbation budget. 

\subsection{Adversarial attacks}
\label{Sec:attacks}
A wide range of gradient-based or optimization-based attacks have been proposed to generate AEs under different types of norms. 
DeepFool~\cite{deepfool} is an $L_2$-norm attack method which performs a minimum amount of perturbations by moving the attacked input sample to its closest decision boundary. Another effective $L_2$-norm attack proposed by Carlini and Wagner (CW attack)~\cite{cwattack} takes a Lagrangian form and uses Adam~\cite{adam} for optimization. 

In this work, we focus on SOTA $L_\infty$ adversarial attacks, which are mostly used according to its consistency with respect to human perception~\cite{PGD}. The conventional $L_\infty$ approaches can be categorized into three categories. The first category is a \textit{one-step gradient-based approach}, such as the fast gradient sign method (FGSM) \cite{goodfellow2014explaining}, which generates an  adversarial example $x^*$ by minimizing the loss $J(h_\theta(x^*), c)$. $J$ is often chosen as the cross-entropy loss and $\epsilon$ is the $L_\infty$ norm bound: 
\begin{align}
   x^* = x - \epsilon \cdot \mathrm{sign}(\nabla_x J(h_\theta(x^*), c)),
\label{Eq:FGSM}
\end{align}
The second category is an \textit{iterative method}. The basic iterative method (BIM) \cite{bim} is an iterative version of FGSM, which iteratively updates perturbations with a smaller step size $\alpha$ and keep the perturbation in the $L_\infty$ norm bound by the projection function $\Pi(\cdot)$:
\begin{align}
   x_{t+1}^* = \Pi_\epsilon( x_t^* - \alpha \cdot \mathrm{sign}(\nabla_x J(h_\theta(x_t^*), c))),~ x_0^* = x,
\label{Eq:BIM}
\end{align}
Different from BIM, another iterative method named projected gradient descent (PGD)~\cite{PGD} uses a random start $x_0 = x + U^d(-\epsilon, \epsilon)$, where $U^d(-\epsilon,\epsilon)$ is the uniform noise between $-\epsilon$ and $\epsilon$, and perturbs the input iteratively by Eq.~(\ref{Eq:BIM}). Furthermore, the momentum iterative method (MIM)~\cite{dong2018boosting} is proposed to improve the transferability by integrating the momentum term into the iterative process. Translation-Invariant Attacks (TIM)~\cite{dong2019evading} boosts BIM by optimizing a perturbation over an ensemble of translated images.

The last category is \textit{optimization-based method}, among which one of the representative methods is the Carlini and Wanger (CW) attack \cite{cwattack}. According to~\cite{PGD}, the $L_\infty$ version of CW attack can be solved by the PGD algorithm using the following loss function:
\begin{align}
   \hat{J} = \max(l_{c}(x_t^*, \theta) - l_{{y_{max} \neq c}}(x_t^*, \theta), -\kappa).
\label{Eq:CW}
\end{align}
where $l_{y_{max} \neq c}$ is the largest logit of the remaining classes except for class $c$; $l_{c}$ is the logit with respect to the target class; and $\kappa$ is the parameter controlling  the confidence.\footnote{We set $\kappa$ to the average difference between the largest logit and the penultimate logits for each dataset.} $\hat{J}$ can better compromise the classifier by making the second largest logit $l_{{y_{max} \neq c}}$ smaller than $l_c - \kappa$~\cite{cwattack}. It is worth noting that since our HFC is an add-on, it works with any of the aforementioned attacking method.

\subsection{Adversarial defenses}

Plenty of \textit{proactive defense} approaches have been proposed to clear the adversarial perturbation, such as network distillation ~\cite{papernot2016distillation}, feature squeezing~\cite{xu2017feature}, input smoothing transformation (\eg autoencoder-based denoising~\cite{liao2018defense}, JPEG compression~\cite{jpeg} and regularization~\cite{ross2017improving}), Parseval network~\cite{cisse2017parseval}, gradient masking~\cite{masking}, randomization~\cite{liu2018towards,dhillon2018stochastic}, radial basis mapping kernel~\cite{taghanaki2019kernelized}, and non-local context encoder~\cite{he2019non}. According to~\cite{dong2019benchmarking}, the PGD-based adversarial (re)training~\cite{goodfellow2014explaining,tramer2017ensemble,PGD} achieves the strongest robustness by augmenting the training set with adversarial examples, but consumes accuracy and training time. However, these defenses can be bypassed either completely or partially by adaptive attacks~\cite{CW_bpda,CW_ten,tramer2020adaptive}. 

Another line of research is \textit{reactive defense}, aiming at detecting AEs with high accuracy~\cite{meng2017magnet,miller2020adversarial,zheng2018robust}. In particular, several emerging works shed light on the intrinsic characteristic in the high dimensional feature subspace~\cite{zheng2018robust,li2017adversarial}. Some of them use learning-based methods (\eg DNN~\cite{metzen2017detecting}, RBF-SVM~\cite{SaftyNet}) to train a decision boundary between adversarial and clean distributions in the feature space. Others use k-nearest neighbors (kNN)~\cite{dubey2019defense,papernot2018deep,cohen2020detecting} based methods, which make prediction according to logits (or classes) of the kNNs in the feature space. Furthermore, anomaly-detection based methods are suitable for detecting AEs too. Feinman \textit{et al.}~\cite{kde} and Li \textit{et al.}~\cite{li2020robust} model the normal distribution with kernel density estimation (KDE) and multivariate Gaussian model (MGM), respectively. Ma \textit{et al.}~\cite{ma2018characterizing} characterize the dimensional properties of the adversarial subspaces by local intrinsic dimensionality (LID). Lee \textit{et al.}~\cite{MAHA} measure the degree of the outlier by a Mahalanobis distance (MAHA) based confidence score. 

Regarding the detectability of medical AEs, Ma \textit{et al.}~\cite{ma2020understanding} evaluate that medical AEs are much easier to detect than natural images (with \textit{100\%} accuracy). A similar conclusion comes from~\cite{li2020robust}, which motivates us to explore the reason behind this phenomenon and evaluate the robustness of those detectors.

\begin{figure}[t]
\begin{center}
  \includegraphics[width=0.65\linewidth]{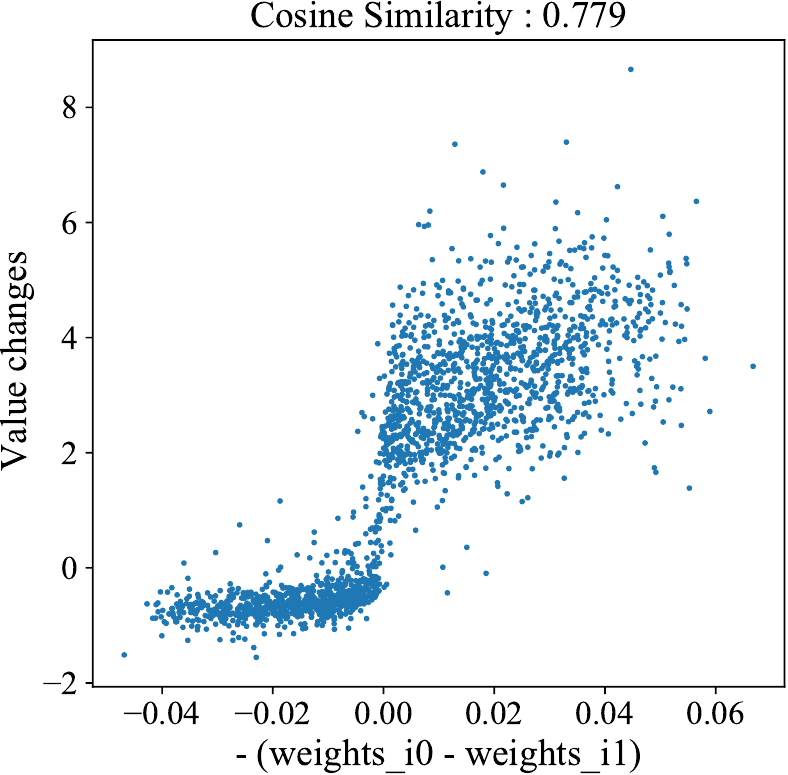}
\end{center}
\vspace{-0.4cm}
  \caption{The similarity between the value changes (under adversarial attack) from the penultimate layer and the difference between $w_{i1}$ and $w_{i0}$. The activation value of the component with a greater difference is increased higher after the attack. }
 \vspace{-0.2cm}
\label{Fig:cos_weight}
\end{figure}

\begin{figure*}[t]
\begin{center}
      \includegraphics[width=0.9\linewidth]{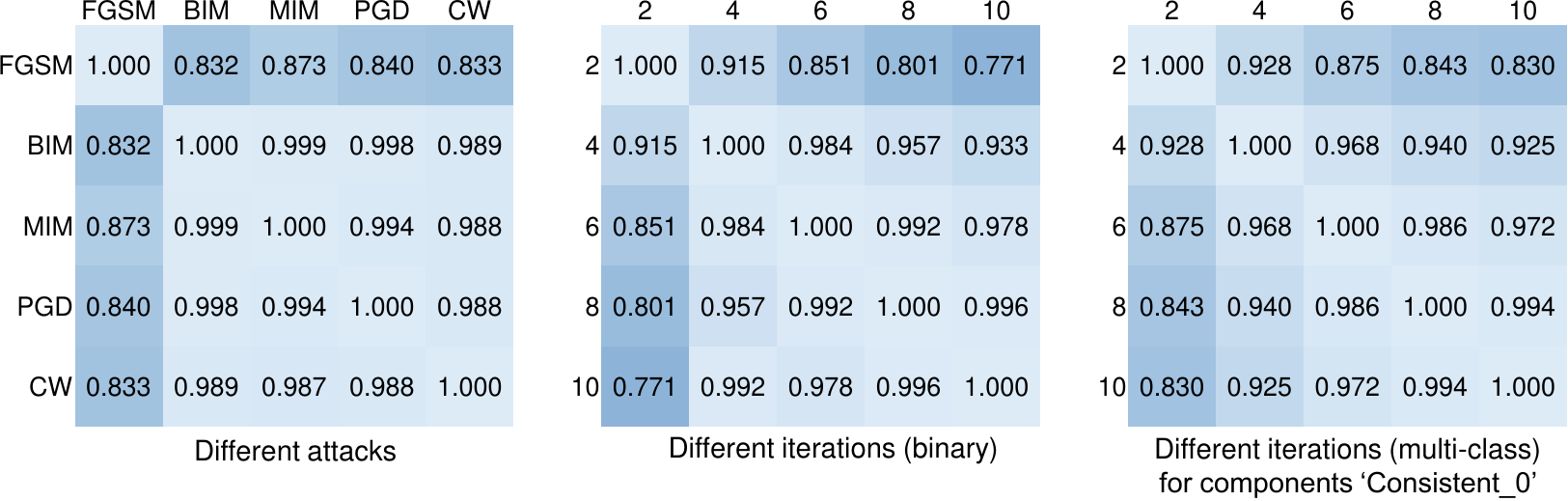}
\end{center}
\vspace{-0.2cm}
   \caption{The similarity of the changes of activation values from the penultimate layer between different attacks and the different iterations from the basic iterative method (BIM) for both binary (on Fundoscopy~\cite{aptos} dataset) and multi-class (on CIFAR-10~\cite{cifar}) classification networks. Here, we choose the  components `Consistent\_0' in the penultimate layer for multi-class setting, whose gradients keep negative in each iteration of BIM. The adversarial attacks are pushed to similar directions between different attacks and iterations.}
\vspace{-0.5cm}
\label{Fig:Similarity}
\end{figure*}

\section{Are Medical AEs Easy to Detect?}



\subsection{Consistency of gradient direction}
\label{Sec:consistent}

To investigate how the vulnerability of feature representations can be exploited by adversarial attacks, we focus on the object function of the final logit output $l_k(x, \theta)$ and its corresponding gradient. In each iteration of the $L_\infty$ approaches introduced in Sec.~\ref{Sec:attacks}, $J$ and $\hat{J}$ simultaneously increase the logit of the target class and decrease the other logits. Therefore, the gradients are towards the similar directions in different iterations of various attacks, which are back-propagated to all of the layers in the diagnosis network according to the chain rule. In this section, we provide empirical and theoretical analyses for both binary and multi-class settings, and discuss the implications in detail.

\subsubsection{Binary classification}

\textbf{Theorem I.} \textit{Consider a binary disease diagnosis network and its representations from the penultimate layer, the directions of the corresponding gradients are fixed during each iteration under adversarial attack.}

\textit{Proof:} For simplicity, we ignore the bias parameters in the penultimate layer, which does not affect the conclusion. Formally, the cross-entropy loss is defined as:
\begin{equation}
    J_{CE} = -\sum_k y_klog(p_k),
\end{equation}
where $p_k$ is computed according to Eq.~(\ref{Eq:inference}).

According to the chain rule, we compute the partial derivative of $J$ with respect to $l_i$:
\begin{equation}
\label{eq:gradient_J_l}
    \nabla_{l_k} J(h_\theta(x)) = p_k - y_k.
\end{equation}

The binary diagnosis network has two classes $k \in [0, 1]$. Hence, 
\begin{equation}
    \begin{split}
        \nabla_{z_i} J(h_\theta(x)) &= \nabla_{l_0} J \times \nabla_{z_i} l_0 + \nabla_{l_1} J \times \nabla_{z_i} l_1 \\
    &= (p_0 - y_0)w_{i0} + (p_1 - y_1)w_{i1}.
    \end{split}
\end{equation}
Without loss of generality, suppose that the goal of targeted adversarial attack is to invert the prediction from 0 to 1, \textit{i.e.}, $y_0=0, y_1=1$. So we can derive the partial derivative on $z_i$:
\begin{equation}
    \begin{split}
        \nabla_{z_i} J(h_\theta(x), y_{1}) &= p_0w_{i0} + (p_1 - 1)w_{i1} \\
        &= (1 - p_1)(w_{i0} - w_{i1}),
    \end{split}
\end{equation}
where $1 - p_1>0$ and $w_{i0} - w_{i1}$ is constant, so the partial derivative on the feature of penultimate layer $z_{i}$ keeps in the same direction.


\textit{Implication.} Accordingly, guided by the gradient, the component $z_i$ with the bigger difference between $w_{i1}$ and $w_{i0}$ increases more drastically under adversarial attack. We plot the similarity between $w_{i0} - w_{i1}$  and the value changes in Fig.~\ref{Fig:cos_weight}. And, the similarities of the value changes of the features in the penultimate layer among different adversarial approaches and different iterations are calculated and presented in Fig.~\ref{Fig:Similarity}. As shown, the BIM attack pushes the features toward similar directions in each iteration. Furthermore, various adversarial attacks push the feature towards a similar direction by increasing the targeted logit and decreasing the left ones, which empirically validates our hypothesis: \textbf{the features are pushed toward a similar direction during different iterations of different attacks}. As a consequence, the adversarial features are pushed to be the outliers and easy to be detected by the reactive defenses.

\subsubsection{Multi-class classification}

\textbf{Theorem II.} \textit{Similar to binary classification, some of the activation values are pushed to outlier, which are guided by the gradients toward similar direction in each iteration.} Differently, the derivative result shows that only the channels ($i$) with the biggest weight $w_{ic}$ (greater than other weights $w_{ik}, k \neq c$) have the negative gradients all the time. We mark these channels as `Consistent\_c'.

\textit{Proof:} The goal is to make the prediction classified to a particular erroneous class $c$. The partial derivative of $J$ with respect to the $i$-th activation of the penultimate layer is computed as:
\begin{equation}
    \nabla_{z_i} J(h_\theta(x)) = \sum_{k=0}^K \nabla_{l_k} J \times \nabla_{z_i} l_k.
\end{equation}
Using Eq.~(\ref{eq:gradient_J_l}), we have:
\begin{equation}
\begin{split}
\label{eq:gradient_J_z}
    \nabla_{z_i} J(h_\theta(x)) = \sum_{k=0}^K(p_k - y_k)w_{ik}.
\end{split}
\end{equation}
For targeted attack, we know that:
\begin{equation}
	y_k = \begin{cases}
	1, &\text{if $k = c$}\\
	0 , &\text{if $k \neq c$}. \\
		   \end{cases}
\end{equation}
Hence, Eq.~(\ref{eq:gradient_J_z}) can be rewritten as:
\begin{equation}
\begin{split}
    \nabla_{z_i} J(h_\theta(x), y_c) &= \sum_{k\neq c}^K p_kw_{ik} + (p_{c} - 1)w_{ic}\\
    &=\sum_{k\neq c}^K p_k(w_{ik} - w_{ic}),
\end{split}
\end{equation}
where $p_k \ge 0$ and $(w_{ik} - w_{ic})$ is constant. Different from the binary setting where all of gradients $\nabla_{z_i} J(h_\theta(x), y_c)$ of the components $z_i$ can keep positive or negative in each iteration, 
we focus on the component $z_i$ of the channels `Consistent\_c' whose $w_{ic}$ is no smaller than the other weights $w_{ik}$:
\begin{equation}
    \forall k \in K, w_{ik} - w_{i\mathrm{c}} \leq 0,
    \label{eq:max_weight}
\end{equation}
where gradient $\nabla_{z_i} J(h_\theta(x), y_{c})$ keeps negative all the time. 

\textit{Implication.} Similar to binary classification, we calculate the changes of the activation values from the  components `Consistent\_0' in the penultimate layer. The results are shown in Fig.~\ref{Fig:Similarity}, which validate Theorem 2: \textbf{the features of these `Consistent\_c' components are pushed toward similar direction during each iteration of the adversarial attack.} Accordingly, these features of components `Consistent\_c' are optimized toward similar direction into abnormal regions and can be easily distinguished by adversarial detectors. 

\begin{figure*}[t]
\begin{center}
   \includegraphics[width=1.0\linewidth]{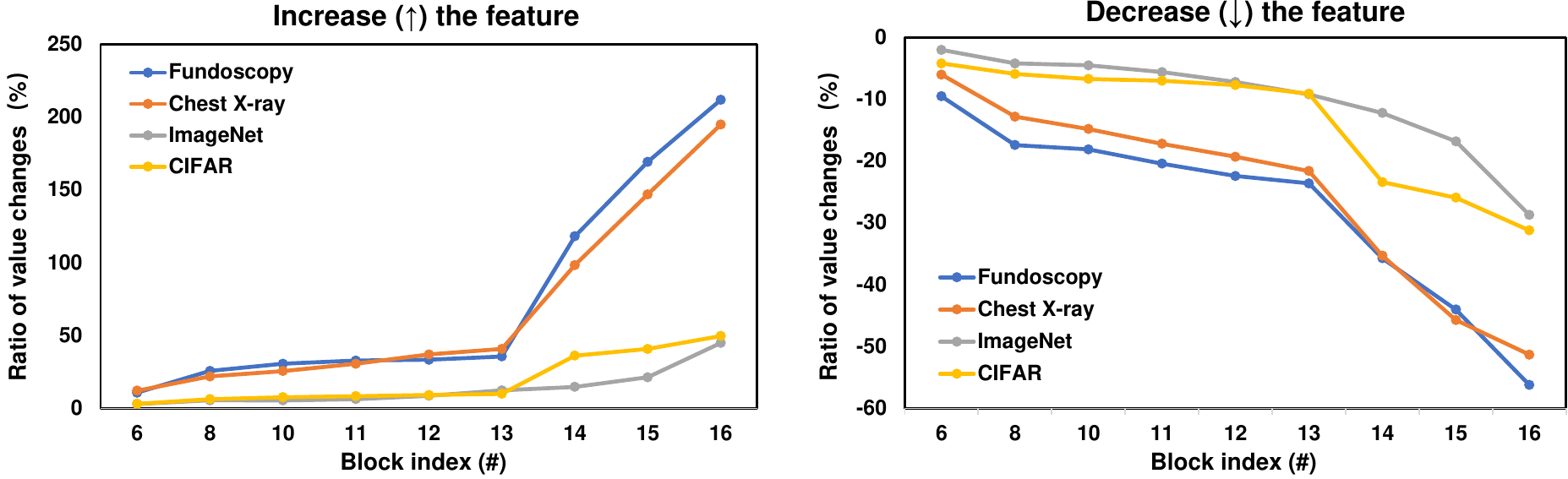}
\end{center}
\vspace{-0.1cm}
   \caption{Comparison of the vulnerability between medical images (Fundoscopy~\cite{aptos} and Chest X-ray~\cite{CXR}) and natural images (CIFAR-10~\cite{cifar} and ImageNet~\cite{imagenet}). We calculate the changed ratios of the activation value from different blocks in ResNet-50 before and after the adversarial attacks. The stress test uses BIM~\cite{bim} with perturbations under the constraint $L_\infty=1/256$, and optimizes $J^*$ for 10 iterations. }
\vspace{-0.1cm}
\label{Fig:Stress_test}
\end{figure*}

\subsection{Vulnerability of representations}
\label{Sec:feature_vulnerable}

We perform a stress test to empirically evaluate the robustness of the features of natural and medical images. In practice, we use adversarial attack to manipulate the deep representations as much as possible. In the experiments, we then use BIM~\cite{bim} to generate adversarial examples, which decrease ($\downarrow$) and increase ($\uparrow$) the features, by optimizing the new loss function $J_\downarrow^* = \mathrm{mean}(f^l(x))$ and $J_\uparrow^* = - \mathrm{mean}(f^l(x))$ in each iteration, respectively, where $f^l(x)$ is the feature from the $l^{th}$ activation layer. The stress test is executed on the medical image dataset (Fundoscopy~\cite{aptos} and Chest X-ray~\cite{CXR}) and natural image dataset (CIFAR-10~\cite{cifar} and ImageNet~\cite{imagenet}). The experimental results shown in Fig.~\ref{Fig:Stress_test} demonstrate that the features of the medical images can be altered more drastically, \emph{i.e.,} \textbf{the medical image representations are much more vulnerable than natural images.} 


\begin{table*}[htbp]
\vspace{-0.2cm}
\caption{The point-wise results of the proposed HFC. The metric scores on the \textit{left} of $\rightarrow$ are the accuracy (\%) of the adversarial detectors, trained on the AEs generated by the corresponding adversarial attacks with the constraint $L_\infty=1/256$. The metrics on the \textit{right} of $\rightarrow$ are the adversarial detection accuracy under our attack ($\downarrow$), which satisfies the same constraint. All of the proposed attacks are under constraint $L_\infty=1/256$ and evaluated on ResNet-50 (2D images) or ResNet-18 (3D volumes). Adv. Acc is the successful rate ($\uparrow$) of HFC to manipulate the prediction of disease diagnosis network. }
\centering
\small
\setlength{\tabcolsep}{1mm}{
\begin{tabular}{l|rrr|rrr|rrr|rrr|rrr|rrr|rrr|rrr}
\bottomrule \hline 
 \multirow{2}{*}{Fundoscopy}  & \multicolumn{6}{c|}{MIM (Adv. Acc=99.5\%)} & \multicolumn{6}{c|}{TIM (Adv. Acc=95.9\%)} & \multicolumn{6}{c|}{PGD (Adv. Acc=99.5\%)} & \multicolumn{6}{c}{CW (Adv. Acc=99.5\%)} \\ \cline{2-25} 
        & \multicolumn{3}{c|}{AUC}         &  \multicolumn{3}{c|}{TPR@90}         & \multicolumn{3}{c|}{AUC}         & \multicolumn{3}{c|}{TPR@90}        & \multicolumn{3}{c|}{AUC}         & \multicolumn{3}{c|}{TPR@90}       & \multicolumn{3}{c|}{AUC}         & \multicolumn{3}{c}{TPR@90}       \\ 
\hline
KD~\cite{kde}        & 98.8 & $\rightarrow$ & 72.0   & 96.3 & $\rightarrow$ & 10.0  & 99.3 & $\rightarrow$ &69.9  & 98.2 & $\rightarrow$ &18.2    & 99.4 & $\rightarrow$ &73.4  & 98.6 & $\rightarrow$ & 13.2  & 99.5 & $\rightarrow$& 74.7  & 99.1 & $\rightarrow$ &19.6 \\
MAHA~\cite{MAHA}      & 100 & $\rightarrow$ & ~7.8     & 100 & $\rightarrow$ & ~0.0      & 100 & $\rightarrow$ & ~8.5   & 100 & $\rightarrow$ & ~0.0     & 100 & $\rightarrow$ & ~4.2    & 100 & $\rightarrow$ & ~0.0      & 99.8 & $\rightarrow$ & 33.0  & 99.5 & $\rightarrow$ & ~0.0    \\
LID~\cite{ma2018characterizing}       & 98.8 & $\rightarrow$ & 67.1   & 99.1 & $\rightarrow$ & 31.5  & 98.0 & $\rightarrow$ & 59.4  & 93.6 & $\rightarrow$ & 17.2   & 99.6 & $\rightarrow$ & 73.2  & 98.6 & $\rightarrow$ & 35.5  & 98.8 & $\rightarrow$ & 73.4  & 97.7 & $\rightarrow$ & 33.3 \\
SVM~\cite{SaftyNet}       & 96.9 & $\rightarrow$ & 27.3   & 99.5 & $\rightarrow$ & 27.3  & 99.6 & $\rightarrow$ & 46.2  & 99.1 & $\rightarrow$ & ~11.0     & 99.8 & $\rightarrow$ & 23.1  & 99.5 & $\rightarrow$ & ~0.0     & 99.8 & $\rightarrow$ & 27.0  & 99.5 & $\rightarrow$ & ~0.0    \\
DNN~\cite{metzen2017detecting}       & 100 & $\rightarrow$ & 31.5    & 100 & $\rightarrow$ & ~0.5   & 100 & $\rightarrow$ & 70.0     & 100 & $\rightarrow$ & ~9.1   & 100 & $\rightarrow$ & 58.6   & 100 & $\rightarrow$ & ~8.2    & 100 & $\rightarrow$ & 62.6   & 100 & $\rightarrow$ & 15.1  \\
BU~\cite{kde}        & 89.9 & $\rightarrow$ & 33.5   & 60.7 & $\rightarrow$ & ~0.0     & 88.5 & $\rightarrow$ & 18.0  & 73.2 & $\rightarrow$ & ~0.0      & 61.9 & $\rightarrow$ & 35.9  & 9.1 & $\rightarrow$ & ~0.0      & 93.0 & $\rightarrow$ & 32.8  & 73.1 & $\rightarrow$ & ~5.0   \\
\hline
 \multirow{2}{*}{Chest X-ray} & \multicolumn{6}{c|}{MIM (Adv. Acc=98.1\%)} & \multicolumn{6}{c|}{TIM (Adv. Acc=85.8\%)} & \multicolumn{6}{c|}{PGD (Adv. Acc=90.9\%)} & \multicolumn{6}{c}{CW (Adv. Acc=98.9\%)} \\ \cline{2-25} 
        & \multicolumn{3}{c|}{AUC}          & \multicolumn{3}{c|}{TPR@90}        & \multicolumn{3}{c|}{AUC}         & \multicolumn{3}{c|}{TPR@90}        & \multicolumn{3}{c|}{AUC}         & \multicolumn{3}{c|}{TPR@90}        & \multicolumn{3}{c|}{AUC}         & \multicolumn{3}{c}{TPR@90}     \\ 
\hline
KD~\cite{kde}                 & 100 & $\rightarrow$ & 67.9   & 100 & $\rightarrow$ & ~7.9  & 99.3 & $\rightarrow$ & 74.6    & 98.8 & $\rightarrow$ & ~11.0  & 100 & $\rightarrow$ & 82.3    & 100 & $\rightarrow$ & 50.5  & 99.2 & $\rightarrow$ & 71.5   & 98.4 & $\rightarrow$ & 15.7    \\
MAHA~\cite{MAHA}               & 100 & $\rightarrow$ & ~0.0    & 100 & $\rightarrow$ & ~0.0  & 100 & $\rightarrow$ & ~5.4   & 100 & $\rightarrow$ & ~0.0    & 100 & $\rightarrow$ & ~0.0  & 100 & $\rightarrow$ & ~0.0     & 100 & $\rightarrow$ & 22.4          & 100 & $\rightarrow$ & ~0.0     \\
LID~\cite{ma2018characterizing}                & 100 & $\rightarrow$ & 47.5      & 100 & $\rightarrow$ & ~2.3   & 99.6 & $\rightarrow$ & 61.1    & 99.7 & $\rightarrow$ & 28.6   & 100 & $\rightarrow$ & 49.1    & 100 & $\rightarrow$ & ~1.5   & 99.2 & $\rightarrow$ & 64.5     & 98.4 & $\rightarrow$ & 14.4    \\
SVM~\cite{SaftyNet}                  & 100 & $\rightarrow$ & 8.9 & 100 & $\rightarrow$ & 46.7 & 98.5 & $\rightarrow$ & 6.9    & 99.6 & $\rightarrow$ & ~2.3     & 100 & $\rightarrow$ & ~5.8    & 100 & $\rightarrow$ & ~0.0     & 100 & $\rightarrow$ & 21.2  & 100 & $\rightarrow$ & ~0.0 \\
DNN~\cite{metzen2017detecting}                & 100 & $\rightarrow$ & 35.5  & 100 & $\rightarrow$ & ~1.0     & 100 & $\rightarrow$ & 55.6    & 100 & $\rightarrow$ & ~0.9   & 100 & $\rightarrow$ & 33.7    & 100 & $\rightarrow$ & ~0.0     & 100 & $\rightarrow$ & 61.6    & 100 & $\rightarrow$ & ~5.2     \\
BU~\cite{kde}                 & 100 & $\rightarrow$ & 15.2    & 100 & $\rightarrow$ & ~0.0    & 94.6 & $\rightarrow$ & 7.1   & 83.3 & $\rightarrow$ & ~0.0    & 49.2 & $\rightarrow$ & 26.2   & 22.7 & $\rightarrow$ & ~0.0    & 98.3 & $\rightarrow$ & 26.2      & 94.8 & $\rightarrow$ & ~0.0     \\
\hline
 \multirow{2}{*}{Brain} & \multicolumn{6}{c|}{MIM (Adv. Acc=100\%)} & \multicolumn{6}{c|}{TIM (Adv. Acc=100\%)} & \multicolumn{6}{c|}{PGD (Adv. Acc=100\%)} & \multicolumn{6}{c}{CW (Adv. Acc=100\%)} \\ \cline{2-25} 
        & \multicolumn{3}{c|}{AUC}          & \multicolumn{3}{c|}{TPR@90}        & \multicolumn{3}{c|}{AUC}         & \multicolumn{3}{c|}{TPR@90}        & \multicolumn{3}{c|}{AUC}         & \multicolumn{3}{c|}{TPR@90}        & \multicolumn{3}{c|}{AUC}         & \multicolumn{3}{c}{TPR@90}     \\ 
\hline
KD~\cite{kde}                 & 100 & $\rightarrow$ & 50.8   & 100 & $\rightarrow$ & 14.5  & 100 & $\rightarrow$ & 56.3   & 100 & $\rightarrow$ & 13.5  & 100 & $\rightarrow$ & 61.5    & 100 & $\rightarrow$ & 23.1  & 100 & $\rightarrow$ & 60.6   & 100 & $\rightarrow$ & 14.3    \\
MAHA~\cite{MAHA}               & 100 & $\rightarrow$ & ~0.0    & 100 & $\rightarrow$ & ~0.0  & 100 & $\rightarrow$ & ~0.0   & 100 & $\rightarrow$ & ~0.0    & 100 & $\rightarrow$ & ~0.0  & 100 & $\rightarrow$ & ~0.0     & 100 & $\rightarrow$ & 0.0          & 100 & $\rightarrow$ & ~0.0     \\
LID~\cite{ma2018characterizing}                & 100 & $\rightarrow$ & 50.3      & 100 & $\rightarrow$ & 10.0   & 100 & $\rightarrow$ & 57.2   & 100 & $\rightarrow$ & 18.8   & 100 & $\rightarrow$ & 49.3   & 100 & $\rightarrow$ & 23.1   & 100& $\rightarrow$ & 58.2     & 100 & $\rightarrow$ & 21.3    \\
SVM~\cite{SaftyNet}                & 100 & $\rightarrow$ & 72.9 & 100 & $\rightarrow$ & 54.3 & 100 & $\rightarrow$ & 66.5  & 100 & $\rightarrow$ & 34.5     & 100 & $\rightarrow$ & 75.5    & 100 & $\rightarrow$ & 51.5     & 100 & $\rightarrow$ & 77.9 & 100 & $\rightarrow$ & 51.0 \\
BU~\cite{kde}                 & 100 & $\rightarrow$ & 45.8    & 100 & $\rightarrow$ & 8.6    & 100 & $\rightarrow$ & 35.4   & 100& $\rightarrow$ & ~7.6    & 100 & $\rightarrow$ & 43.7   & 100 & $\rightarrow$ & 1.5   & 100& $\rightarrow$ & 38.5      & 100 & $\rightarrow$ & 2.7     \\
\hline \toprule 
\end{tabular}
}
\vspace{-0.5cm}
\label{Table:pointwise}
\end{table*}

\section{Adversarial attack with a hierarchical feature constraint}
\label{Sec:Method}

In this section, we demonstrate a way of hiding the adversarial representation in the target feature distribution. Our intuition is to derive a term that measures the distance from the adversarial representation to this distribution; therefore, the adversarial representation can be pushed toward such a distribution on the shortest path by directly minimizing the defined term during the process of stochastic gradient descent in each iteration of adversarial attack.

\textbf{Modeling the target feature distribution:} 
First, the feature distribution of the target class is modeled using a Gaussian mixture model (GMM) as follows:
\begin{equation}
    p(f_\theta^l(x)) = \sum_{m=1}^{M}\pi_m \mathcal{N}(f_\theta^l(x)|\mu_{m}, \Sigma_{m}),
\end{equation}
where $p$ is the probability density of sample $x$ with respect to the target class $c$; $f_\theta^l(\cdot)$ denotes the mapping function, \eg the deep representation of the $l^{th}$ activation layer with parameters $\theta$; $\pi_m$ is the mixture coefficient subject to $\sum_{m=1}^{M}\pi_m=1$; and $\mu_m$ and $\Sigma_m$ are the mean and covariance matrix of the $m$-th Gaussian component in the mixture model. We train these parameters by the expectation-maximization (EM) algorithm \cite{EM} on the training images of the target class $c$.

For a given input $x$, 
we separately compute the log-likelihood of an adversarial feature relative to each component, and choose the most probable Gaussian component:
\begin{equation}
m' = \mathop{\arg\max}\limits_m ~ \ln(\pi_m \mathcal{N}(f_\theta^l(x)|\mu_{m}, \Sigma_{m})). \label{eq:k'}
\end{equation}
Then, the log-likelihood of this chosen component $m'$ is maximized to hide the adversarial representation.

\renewcommand{\algorithmicrequire}{ \textbf{Input:}} 
\renewcommand{\algorithmicensure}{ \textbf{Output:}} 

\begin{algorithm}[t]
\caption{BIM attack with hierarchical feature constraint}
\label{algo1}
\begin{algorithmic}[1]
\REQUIRE~~\\
Training set $(x^i, y^i)\in \mathcal{D}$, an input image $x$, target class $c$, a DNN model $h$ with $L$ feature layers $\{f_\theta^l;~l=1:L\}$, the maximum number of iterations $T$, and a step size $\alpha$

\ENSURE ~~\\
Adversarial image $x^*$ with $||x^* - x||_\infty \leq \epsilon$

\FOR{each DNN's layer $l=1$ to $L$}
\STATE Initialize mean $\mu_{m}^l$ and covariance $\Sigma_{m}^l$ of each component, where $m \in \{1, ..., M\}$
\STATE Train the GMM using the EM algorithm with $f_\theta^l(x^i)$,  where $x^i \in \mathcal{D}$ with $y^i = c$
\ENDFOR
\STATE $x_0^* = x$
\FOR{$t=1$ to $T$}
\STATE $J_{final} =  J_{\mathrm{HFC}} + \dot{J}$,
\STATE $x_{t+1}^* = \mathrm{clip}_\epsilon( x_t^* - \alpha \cdot \mathrm{sign}(\nabla_x (J_\mathrm{HFC} + \dot{J})))$
\ENDFOR
\RETURN $x^*$
\end{algorithmic}
\end{algorithm}

\textbf{Hierarchical feature constraint:} To avoid being detected by outlier detectors, we add the constraint $\ln(\pi_m \mathcal{N}(f_\theta^l(x)|\mu_{m'}, \Sigma_{m'}))$, ignoring the constant terms, to all DNN layers. The hierarchical feature constraint induces a loss ($J_{\mathrm{HFC}}$) that can be formulated as:
\begin{equation}
\label{eq11}
    J_{\mathrm{HFC}} =  \sum^L_{l=1} \frac{\lambda^l}{2}(f_\theta^l(x) - \mu^l_{m'(l)})^\top (\Sigma_{m'(l)}^{l})^{-1} (f_\theta^l(x) - \mu^l_{m'(l)}),
\end{equation}
where $\lambda^l$ is a weighting factor that controls the contribution of constraint in layer $l$, $m'(l)$ is the chosen mixture component for layer $l$, and $L$ is the total number of layers.

Algorithm~\ref{algo1} shows the pseudo-code for the BIM attack with hierarchical feature constraint. 
Given an input image $x$, the goal of our HFC is to find an adversarial example that can be misclassified to the target class $c$, while keeping the deep representation close to the feature distribution of normal samples. Here, we focus on the AEs with the $L_\infty$ constraint. Concretely, we first model the target hierarchical features of the training data with GMM,
and then extend the attacking process of BIM by replacing the original loss function $J$ in Eq.~(\ref{Eq:BIM}) with:
\begin{equation}
    J_{\mathrm{final}} = J_{\mathrm{HFC}} + \dot{J},
\label{Eq:final}
\end{equation}
where $\dot{J}$ is the classification loss used in original attack (e.g., Eq.~(\ref{Eq:CW}) in CW attack and cross-entropy loss $J$ in PGD, BIM, MIM, etc) and $J_{\mathrm{HFC}}$ is the HFC loss term. Furthermore, we can also easily apply the HFC term to other advanced attacks (e.g., MIM, TIM, etc) by \textit{simply adding HFC to the original loss function}.

\begin{figure*}[h]
\begin{center}
\vspace{-8mm}
   \includegraphics[width=0.95\linewidth]{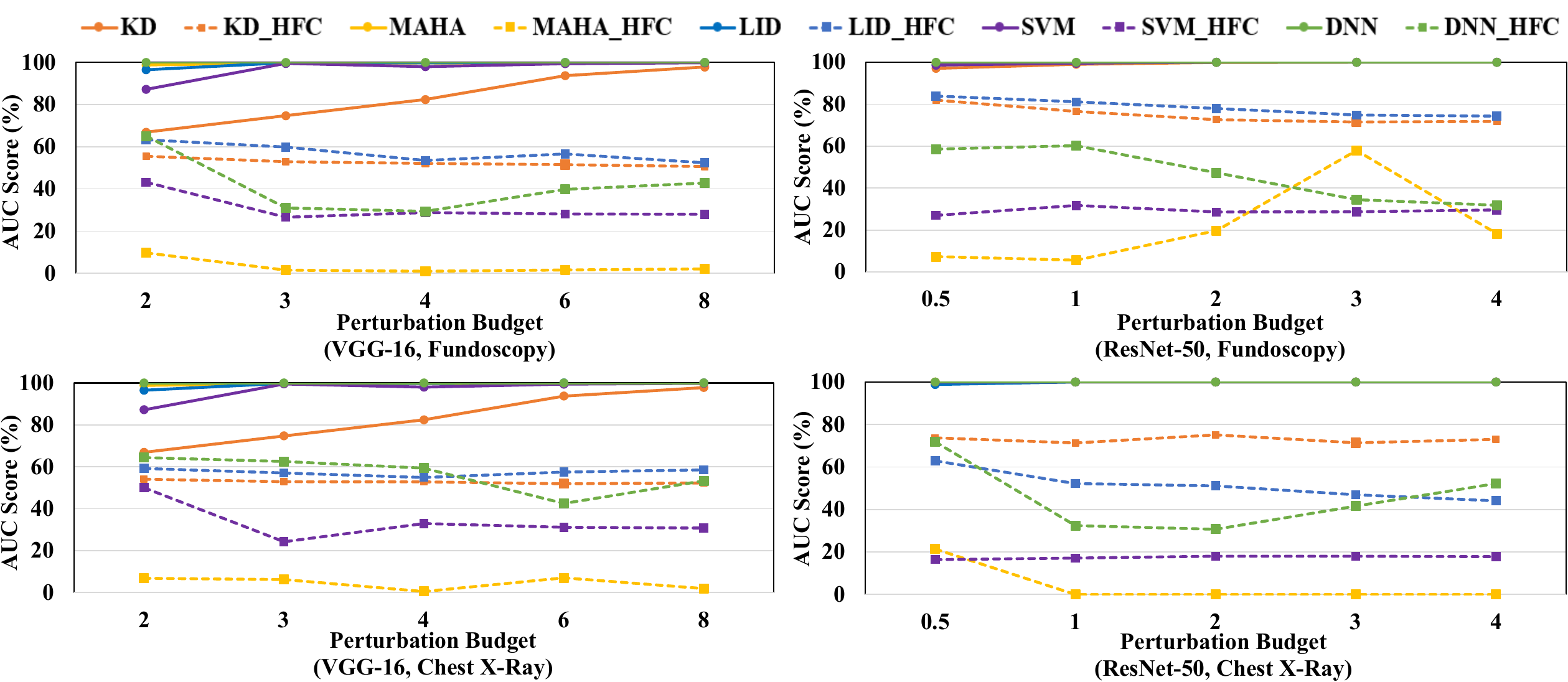}
\end{center}
   \caption{The AUC scores in the solid lines are the performances of the adversarial detectors trained on AEs generated by BIM, where most of the scores are around 100.0\%, resulting in overlapped solid lines at the top of the charts. The dotted lines denote the adversarial detection accuracy under our proposed attack.  The performance drop of each adversarial detector can be observed from the pair of solid and dotted line of corresponding color. Our method can bypass an array of adversarial detectors as indicated by a drastic performance drop from solid line to dotted line on different backbones (left for VGG-16 and right for ResNet-50), datasets (top for Fundoscopy and bottom for Chest X-Ray), and perturbation budgets ($L_\infty/256$ is indicated by the horizontal axis). }
\label{Fig:Pertubation}
\vspace{-5mm}
\end{figure*}

\subsection{HFC for anomaly detection}

Interestingly, we find that HFC has the potential to detect anomaly signals. For the input $x$, we compute the HFC term of the feature in the last layer $n$ \textit{as the image-level anomaly score $S_{\mathrm{HFC}}$} as follows:

\begin{align}
\vspace{-10mm}
    m' &= \mathop{\arg\max}\limits_m ~ \ln(\pi_m \mathcal{N}(f_\theta^n(x)|\mu_{m}, \Sigma_{m})). \label{eq:k} \\
    S_{\mathrm{HFC}} &=  (f_\theta^n(x) - \mu^i_{m'(i)})^\top (\Sigma_{m'(n)}^{n})^{-1} (f_\theta^n(x) - \mu^n_{m'(n)})
\label{Eq:anomaly}
\end{align}
where $f_\theta^n(x)$ is the feature of the last layer of input $x$; $\mu_m$ and $\Sigma_m$ are the mean and covariance matrix of the $m$-th component of GMM for the last layer; $m'$ is the component of the GMM model with the minimum log-likelihood. 

\section{Experiments}


\subsection{Setup}
\label{Sec:setup}

\textbf{Evaluation process and metrics.} The targeted adversarial example (AE) first manipulates the results of the disease classifier $h$, and we report the \textit{Adversarial accuracy (Adv. Acc)} representing the success rate of AE in manipulating $h$. Then, the reactive defense $d$ is supposed to identify the AE that succeeds in attacking the classifier $h$. We report \textit{True positive rate at 90\% true negative rate (TPR@90):} 10\% of the normal samples are dropped by the defense $d$ to reject more AEs; and \textit{Area under curve (AUC) score} of defense $d$ for detecting AEs. A strong attacker should manipulate classifier $h$ (higher Adv. Acc) and compromise the reactive defense $d$ (lower TPR@90 and AUC) simultaneously.

\textbf{Assumptions of the attackers.} In the white-box setting, the attackers are aware of the parameters of the disease classifier $h$, but not aware of any specific strategy or parameters of the deployed reactive defense $d$. While in the semi-white-box setting (in Section V.E), the attackers are only aware of the type of the disease classifier $h$, but not aware of the parameters of $h$ as well as the specifics of the deployed reactive defense $d$. For the computational capabilities, we assume that the attacker has a GPU to generate AEs using HFC.

\textbf{Datasets.} We choose two public datasets on typical medical classification tasks with 2D images, following~\cite{ma2020understanding,finlayson2018adversarial}. (\romannumeral1) Kaggle \textbf{Fundoscopy}~\cite{aptos} dataset on the diabetic retinopathy (DR) diagnosis task, consisting of 3,663 high-resolution fundus images. Each image is labeled to one of the five levels from ‘No DR’ to ‘mid/moderate/severe/proliferate DR’. Following \cite{ma2020understanding,finlayson2018adversarial}, we conduct a binary classification experiment, which considers all fundus images labeled with DR as the same class.  (\romannumeral2) Kaggle\footnote{Kaggle Chest X-ray Images (Pneumonia): \url{https://www.kaggle.com/paultimothymooney/chest-xray-pneumonia}} dataset on the pneumonia diagnosis task, which contains 5,863 \textbf{Chest X-ray} images labeled with  `Normal' or `Pneumonia' . 

Furthermore, we evaluate HFC on a 3D medical volume dataset: Brain hemorrhage (\textbf{Brain}), which contains 1,486 brain CT volumes collected from our collaborative hospital with Institutional Review Boards (IRB) approval. The volumes are labeled according to the pathological causes of cerebral hemorrhage: aneurysm, arteriovenous malformation, moyamoya
disease, and hypertension. Each CT volume is standardized to 30$\times$270$\times$270 voxels. We randomly crop patches with a size of 30$\times$128$\times$128 and normalize them to [0,1] during training.

Following~\cite{ma2020understanding,ma2018characterizing}, each dataset is split into three subsets: \textit{Train}, \textit{AdvTrain} and \textit{AdvTest}. Also, we randomly select 80\% of the images as \textit{Train set} to train the DNN classifier, and the left samples are treated as the \textit{Test set}. We discard the misclassified test samples (by the diagnosis network), dropping 7, 101, 50 samples on Fundoscopy, Chest X-ray, and Brain datasets, respectively. Then the
70\% of the samples (\textit{AdvTrain}) in the \textit{Test set} are used to train the adversarial detectors, and the left 30\% images (\textit{AdvTest}) are used to evaluate their effectiveness. 

\textbf{DNN classifiers.} For 2D medical images (Fundoscopy and Chest X-ray), the ResNet-50 \cite{resnet} and VGG-16 \cite{VGG} models pretrained on ImageNet~\cite{imagenet} are chosen as classifier. All images are resized to 256$\times$256$\times$3 and normalized to [0,1]. The models are trained with augmented data using random crop and horizontal flip. For 3D medical volumes, we choose ResNet-18~\cite{resnet} as our backbone by replacing the 2D convolution kernels with 3D kernels. 

\textbf{Adversarial attacks and detectors.} Following~\cite{ma2020understanding}, we choose MIM, BIM, PGD, TIM~\cite{dong2019evading} and CW to attack our models.
For the adversarial detectors, we choose the following reactive defenses:
\begin{itemize}
    \item \textit{Kernel density (KD)}~\cite{kde}. Given a sample $x$ of class $k$, and a set of training samples from the same class $X_k$, the KD score of $x$ can be estimated by:
\begin{equation}
    \mathrm{KD}(x) = \sum_{x'\in X_k} F(z(x), z(x')),
\label{Eq:KD}
\end{equation}
where $z(\cdot)$ is the activation vector of the penultimate layer, $F(\cdot, \cdot)$ is the kernel function, often chosen as a Gaussian kernel. 

    \item \textit{Bayesian uncertainty (BU)}~\cite{kde}. Given a sample $x$, the BU score of $x$ can be estimated by:
\begin{equation}
    \mathrm{BU}(x) = \frac{1}{N}\sum_{n=1}^T {l_n}^Tl_n-(\frac{1}{N}\sum_{n=1}^T {l_n})^T(\frac{1}{T}\sum_{n=1}^T {l_n)},
\end{equation}
where $\{{l_1}, ..., {l_N}\}$ are the stochastic logits predictions calculated when randomly dropping out 30\% features of the penultimate layer. Here, we set $N = 50$.

    \item \textit{Local intrinsic dimensionality (LID)}~\cite{ma2018characterizing} describes the rate of expansion in the number of data objects as the distance from the reference sample increases. Specifically, given a sample $x$, LID makes use of its distances to the first $N$ nearest neighbors:
\begin{equation}
    \mathrm{LID}(f_l(x)) = -(\frac{1}{N}\sum_{i=n}^N log\frac{r_n(f_l(x))}{r_N(f_l(x))})^{-1},
\label{Eq:LID}
\end{equation}
where $f_l(x)$ is the activation values from the $l^{th}$ intermediate layer; $r_n(f_l(x))$ is the Euclidean distance between $f_l(x)$ and its $n^{th}$ nearest neighbor. LID is computed on each activation layer of the diagnosis network. 

\item \textit{Mahalanobis distance-based confidence score (MAHA)}~\cite{MAHA} utilizes the Mahalanobis distance-based metric to measure the outlier degree of the features. Specifically, we first compute the empirical mean $\mu_l$ and covariance $\Sigma_l$ of the activations for each layer $l$ of the training samples. Then, we compute the MAHA as:
\begin{equation}
    \mathrm{MAHA}(f_l(x)) = (f_l(x) - \mu_l)^\top\Sigma_l^{-1}(f_l(x) - \mu_l),
\label{Eq:maha}
\end{equation}
where $f_l(\cdot)$ is the feature in the $l^{th}$  activation layer of DNN. MAHA is computed on each layer  of the diagnosis network. 

\item \textit{RBF-SVM (SVM)}~\cite{SaftyNet} is a learning-based adversarial detector using RBF-SVM trained on the normal and adversarial features from the penultimate layer.

\item \textit{Deep neural network (DNN)}~\cite{metzen2017detecting} is a learning-based adversarial detector using DNN trained on the normal and adversarial features. We train DNN on each activation layer of the diagnosis network and ensemble these networks by summing up their logits.

\end{itemize}

\begin{table}[t]
\caption{Comparison of other adaptive attacks aiming at manipulating representations. The AEs are generated on ResNet-50 under the constraint $L_\infty=1/256$ using BIM. We choose AUC scores (\%)  and "Adv. Acc" (\%, $\uparrow$) as metrics.}
\centering
\small
\scalebox{0.9}{
\begin{tabular}{l|rrrrrr}
\bottomrule \hline
Fundoscopy & KD   & MAHA & LID  & SVM  & DNN  & Adv. Acc \\
\hline
Random~\cite{feature_iclr}     & 75.1 & 86.1 & 91.7 & 48.2 & 93.7 & \textbf{100.0}      \\
Closest~\cite{feature_iclr}    & 77.0   & 64.0   & 91.0 & \textbf{13.0}   & 79.3 & 81.5     \\
KDE~\cite{CW_ten}        & \textbf{51.6} & 86.5 & 90.9 & 45.3 & 95.0   & \textbf{100.0}      \\
LID~\cite{CW_bpda}        & 87.6 & 85.4 & 93.4 & 61.2 & 96.2 & 95.9    \\
HFC   & 74.2 & \textbf{6.4}  & \textbf{78.3} & 28.6 & \textbf{60.0} & 99.5      \\
\hline
Chest X-ray & KD   & MAHA & LID  & SVM  & DNN & Adv. Acc \\
\hline
Random~\cite{feature_iclr}         & 77.0   & 64.0   & 91.0 & 13.0   & 79.3 & 94.8    \\
Closest~\cite{feature_iclr} & 80.1 & 38.3 & 71.3 & \textbf{9.0}  & 87.7 & 53.1    \\
KDE~\cite{CW_ten}                       & \textbf{58.2} & 66.9 & 71.7 & 15.3 & 95.6 & \textbf{100.0}     \\
LID~\cite{CW_bpda}                      & 84.0 & 66.6 & 77.1 & 28.6 & 96.6 & 70.9  \\
HFC                  & 70.8 & \textbf{0.0} & \textbf{53.6} & 16.7 & \textbf{32.6} & 95.8     \\
\hline \toprule
\end{tabular}
}
\label{Table:sota}
\vspace{-5mm}
\end{table}

The parameters for KD, LID, BU, and MAHA are set according to the original papers. We compute the scores of LID and MAHA for all of the activation layers and train a logistic regression classifier \cite{MAHA,ma2018characterizing}. We implement KD, LID, BU, and MAHA using the official code.\footnote{
KD and BU, \url{https://github.com/rfeinman/detecting-adversarial-samples}\\
LID and MAHA, \url{https://github.com/pokaxpoka/deep_Mahalanobis_detector}}

\textbf{Hyperparameters.} We set $M = 16$ and $64$ for Chest X-ray  and Fundoscopy datasets, respectively. We compute the mean value of the features of each channel in the $l^{th}$ activation layer separately. We set $\lambda_l$ to $\frac{1}{C_l}$, where $C_l$ is the number of channels. As a small adversarial perturbation in medical images increases the loss drastically~\cite{ma2020understanding}, we set a tiny step size $\alpha = 0.02/256$ and the number of iteration $T$ to $2\epsilon / \alpha$.

\subsection{Bypassing adversarial detectors}
\label{Sec:experiments_main}

\begin{figure}[t]
\begin{center}
\vspace{-5mm}
   \includegraphics[width=1\linewidth]{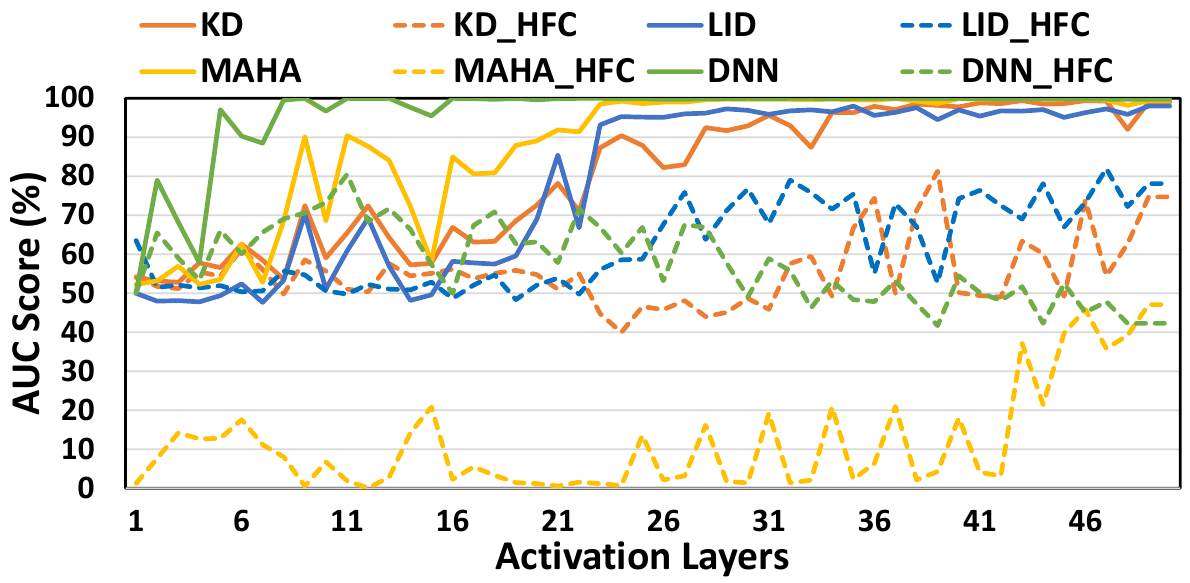}
\end{center}
   \caption{The AUC scores in the solid lines are the performances of the adversarial detectors for each activation layer (from ResNet-50 on Fundoscopy). Our HFC compromises the detection accuracy (the gaps between solid lines and the corresponding dotted lines) at each layer.}
\label{Fig:Layers}
\end{figure}

\begin{figure}[t]
\begin{center}
   \includegraphics[width=1\linewidth]{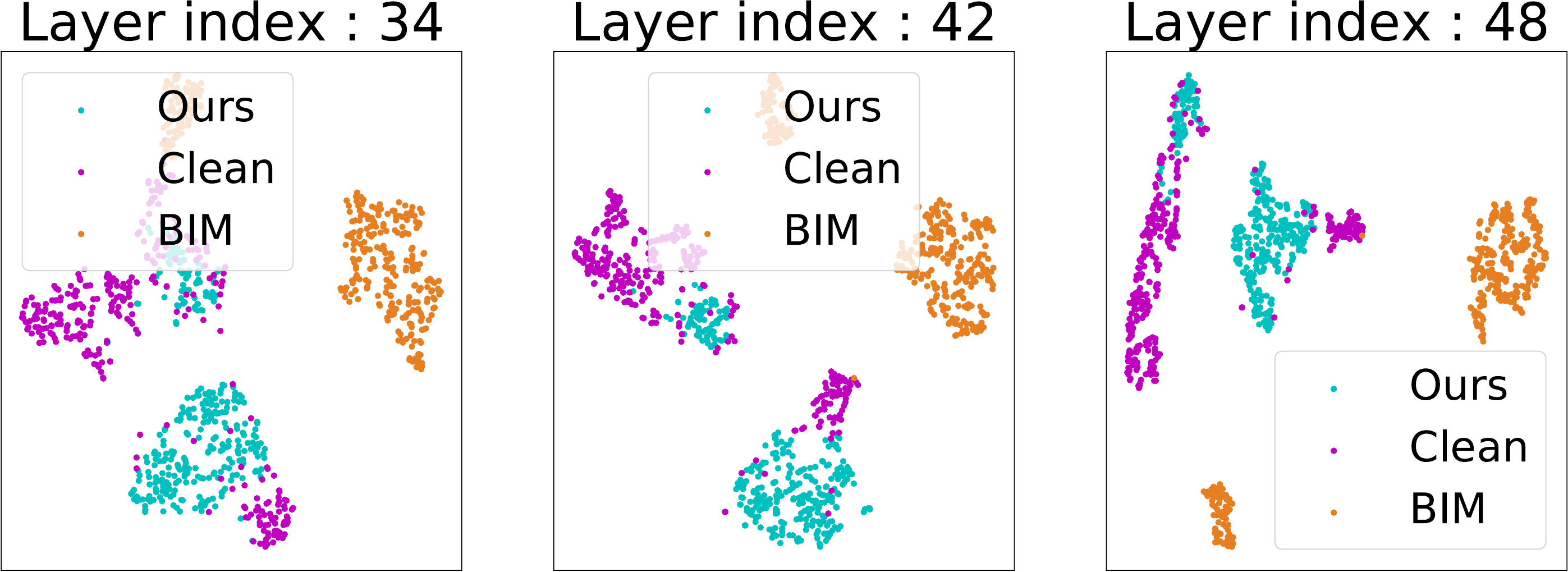}
\end{center}
   \caption{The 2D t-SNE visualization of the features of clean sample and adversarial samples generated from BIM and our method HFC. Features are extracted from ResNet-50 on Chest X-ray. The adversarial examples generated by our method are surrounded by the clean data in the feature space.}
\label{Fig:tsne}
\vspace{-5mm}
\end{figure}

We compare the performances of the existing adversarial attacks \textit{with} and \textit{without} HFC on different datasets, perturbation constraints, and DNN classifiers. The 2D t-SNE in Fig.~\ref{Fig:tsne} illustrates that HFC successfully moves the adversarial features from outlier (orange) regions to the normal regions (cyan) where target features (purple) reside. As shown in Fig.~\ref{Fig:Layers},  the SOTA detectors from all of the activation layers in the diagnosis network are compromised by HFC, \eg the AUC scores of KD and LID are compromised from nearly 100\% to less than 80\%. Moreover, HFC can decrease the MAHA scores to less than clean samples (as the AUC scores decreased to $<$ 20\%) by pushing adversarial features closer to the empirical mean $\mu_l$ in Eq.~(\ref{Eq:maha}).

Quantitative results in Table~\ref{Table:pointwise} justify that four widely-used adversarial attacks are boosted by HFC to bypass an array of SOTA detectors simultaneously with a high accuracy, under the small perturbation constraint $L_\infty=1/256$. On the three datasets covering both 2D and 3D modalities, binary and multi-class settings, most AUC scores of the detectors are decreased to below 80\%. 
Furthermore, when we increase the perturbation budget, the HFC is provided more room to manipulate the representations, compromising the detectors more drastically (as the dotted lines shown in Fig.~\ref{Fig:Pertubation} decrease). Please note that, as different detectors computes anomaly scores in different ways, HFC may not make sure the anomaly scores of AEs are lower than clean samples. For the detectors with AUC scores $<$ 0.5, the decreasing trend may be irregular.

\subsection{Comparison of other adaptive attacks}
\label{Sec:sota}

We compare the effectiveness of our HFC with other competitive adaptive attacks, designed to manipulate the features and bypass the detectors: 1) Generate AEs with internal features similar to a randomly selected guide image \cite{feature_iclr}; 2) Instead of random sampling, select a guide image with its feature closest to the input~\cite{feature_iclr}; 3) Minimize cross-entropy and the loss term of KDE (Eq.~(\ref{Eq:KD})) at the same time~\cite{CW_ten}; 4) Minimize cross-entropy and the loss term of LID (Eq.~(\ref{Eq:LID})) at the same time \cite{CW_bpda}. The results in Table~\ref{Table:sota} show that, all attacks trying to manipulate the adversarial features can successfully weaken the ability of the SOTA detectors. Meanwhile, under the strict constraint ($L_\infty=1/256$), our proposed HFC greatly outperforms other strong adaptive attacks, especially on compromising MAHA, LID and DNN.

\begin{table}[t]
\caption{The AUC (\%) on Fundoscopy dataset of different adversarial detectors on adversarial examples generated by the proposed method with different numbers (M) of GMM components to model the clean data distribution. Smaller numbers indicate better performance.}
\centering
\small
\scalebox{0.9}{
\begin{tabular}{l|rrrrrrrr}
\bottomrule \hline
  $M$  & 1    & 2    & 4    & 8    & 16   & 32   & 64   & 128  \\
\hline
KD  & 75.8 & 78.5 & 77.9 & 77.8 & 73.6 & 73.3 & 74.2 & 74.4 \\
MAHA & 5.9 & 5.6 & 6.2 & 7.1 & 8.2 & 8.6 & 7.6 & 5.9 \\
LID & 81.7 & 82.4 & 83.0 & 84.2 & 83.2 & 80.7 & 78.3 & 78.7 \\
SVM & 52.3 & 45.7 & 40.4 & 34.0 & 31.4 & 34.5 & 35.5 & 40.3 \\
DNN & 61.4 & 60.7 & 62.6 & 64.3 & 64.4 & 63.5 & 63.4 & 64.9 \\
BU & 55.5 & 52.2 & 49.7 & 45.2 & 43.3 & 42.6 & 43.8 & 42.6 \\
\hline
Average & 55.4 & 54.2 & 53.3 & 57.8 & 50.7 & 50.5  & \textbf{50.4} & 51.3\\
Maximum & 81.7 & 82.4 & 83.0 & 84.2 & 83.2 & 80.7 & \textbf{78.3} & 78.7\\
\hline \toprule
\end{tabular}
}
\vspace{-0.4cm}
\label{Table:m}
\end{table}

\subsection{Hyperparameter analysis}
\label{Sec:ablation}

We model the ResNet-50 feature distributions of normal samples on the Fundoscopy dataset by GMM with different numbers of components and evaluate the corresponding attacking performances. As shown in Table~\ref{Table:m}, all of the attacks can stably compromise the detectors, while setting $M=64$ can slightly improve the performance. We speculate that accommodating an excessive number of components within the constrained size of clean samples could pose challenges to the proper fitting of GMMs. Conversely, an insufficient number of components produces a rough fit of the underlying distribution, which might impede the shortest path convergence to push the adversarial feature to the clean side.

\begin{figure*}[ht]
\begin{center}
   \includegraphics[width=0.9\linewidth]{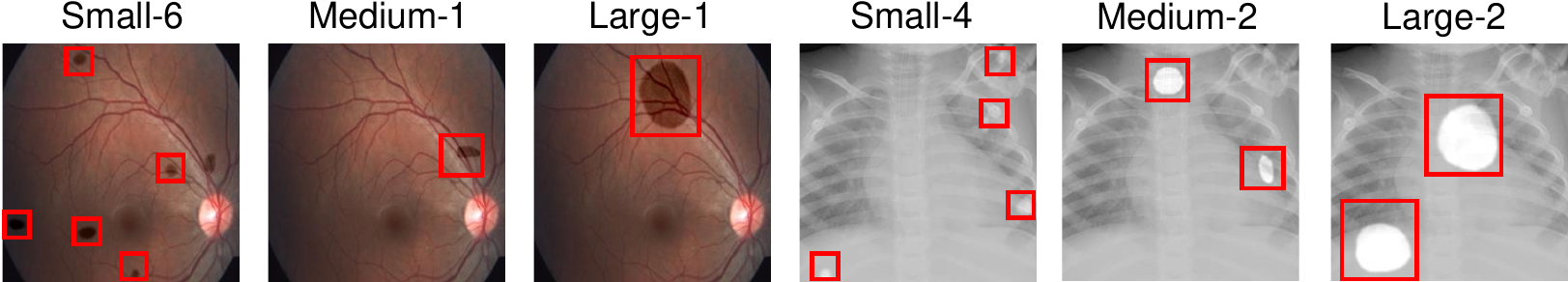}
\vspace{-0.2cm}
\end{center}
   \caption{The visualization of synthesized OOD images. We mark the OOD signals with red boxes, which are ``lesions'' generated by random shapes and textures~\cite{hu2023label,yao2021label}. ``Tiny, Small, Medium, and Large'' are the sizes of lesions, while the numbers are the quantities of the lesions.}
\label{Fig:ood}
\vspace{-0.1cm}
\end{figure*}

\begin{table*}[ht]
\caption{The performance of BIM under the semi-white-box setting \textit{w/} and \textit{w/o} HFC.   The  AEs  are crafted under constraint $L_\infty=4/256$. AUC scores (\%, $\downarrow$) and "Adv. Acc" (\%, $\uparrow$) are used as metric.}
\centering
\begin{tabular}{l|l|rrrrrr|rrrrrr}
\bottomrule \hline
 \multirow{2}{*}{DNN Model} &  \multirow{2}{*}{Attacks} &  \multicolumn{6}{c|}{Fundoscopy~\cite{aptos}} &  \multicolumn{6}{c}{Chest X-ray~\cite{CXR}}      \\ \cline{3-14}
 & & KD   & MAHA   & LID   & SVM   & DNN  & Adv. Acc & KD    & MAHA   & LID   & SVM   & DNN     & Adv. Acc       \\
\hline
 \multirow{2}{*}{ResNet-50} & BIM   & 98.7                 & 100.0                 & 99.5                 & 92.1                 & 100.0                  & 83.2       & 92.6                 & 95.4                 & 79.5                 & 24.1                 & 99.5                 & 85.8           \\
 & BIM w/ HFC  & 78.0                   & 9.1                  & 68.0                   & 16.9                 & 43.8                 & 68.2           & 89.4                 & 85.9                 & 71.5                 & 11.4                 & 79.5                 & 76.4             \\ \cline{3-14}
  \cline{1-2} \multirow{2}{*}{VGG-16} & BIM     & 72.3                 & 89.6                 & 81.5                 & 48.1                 & 95.2                 & 88.6    & 46.4                 & 89.2                 & 81.3                 & 73.6                 & 98.5                 & 98.3               \\
& BIM w/ HFC      & 50.9                 & 18.2                 & 64.6                 & 28.8                 & 16.7                 & 73.2             & 34.6                 & 8.6                  & 49.4                 & 41.3                 & 69.5                 & 88.9             \\
\hline \toprule
\end{tabular}
\vspace{-0.4cm}
\label{Table:gray-box}
\end{table*}

\subsection{Semi-white-box attack}
\label{Sec:gray}

We evaluate the proposed HFC in a more difficult scenario: only the architecture information of DNN models is available. The attacker aims at confusing the victim model and bypassing its adversarial detectors simultaneously, without accessing the parameters of the victim model. In the experiments, we train substitute diagnosis networks and detectors using the same training data and generate AEs using BIM~\cite{bim} \textit{w/} and \textit{w/o} HFC. As shown in Table~\ref{Table:gray-box}, our HFC is able to help BIM bypass most of the detectors and control the prediction of the victim model at the same time, posing more disturbing concerns to the safety of DNNs for further clinical application.

\begin{table}[t]
\caption{The comparison between the attacks in the medical and natural image domains using HFC. The adversarial attacks are deployed on ResNet-50. AUC scores (\%, $\downarrow$) are used as metrics for adversarial detectors.}
\centering
\small
\begin{tabular}{l|rrrrrrr}
\bottomrule \hline
CIFAR-10                 & KD                   & MAHA                 & LID                  & SVM                  & DNN                     \\
\hline
BIM$|L_\infty$=8/256   & 76.4                 & 91.4                  & 80.8                 & 96.9                 & 99.8                \\
Ours$|L_\infty$=8/256   & 30.1                   & 82.7                  & 58.7                   & 87.9                 & 87.3                \\
\hline
CIFAR-10                 & KD                   & MAHA                 & LID                  & SVM                  & DNN                        \\
\hline
BIM$|L_\infty$=16/256   & 87.5                 & 99.0                  & 90.3                 & 99.0                 & 100.0                            \\
Ours$|L_\infty$=16/256   & 24.0                   & 76.4                  & 54.1                   & 85.8                 & 80.7                \\
\hline
Chest X-ray                & KD                   & MAHA                 & LID                  & SVM                  & DNN                            \\
\hline
BIM$|L_\infty$=1/256   & 100.0                 & 100.0                  & 100.0             & 100.0                & 100.0                             \\
Ours$|L_\infty$=1/256   & 71.3                   & 0.0                & 48.6                 & 16.7                & 31.8               \\
\hline \toprule 
\end{tabular}
\label{Table:Cifar}
\end{table}

\subsection{Comparing Chest X-ray and CIFAR-10 attacks}

We also evaluate the Chest X-ray performance of the adversarial detectors and HFC under constraints $L_\infty = 8/256$ and $16/256$ on CIFAR-10~\cite{cifar}, a natural image dataset. The Adv Acc. will be less than 95\% for Chest X-ray and CIFAR-10 when the perturbation is smaller than $0.5/256, 8/256$, respectively. As shown in Table~\ref{Table:Cifar}, a greater perturbation budget changes the adversarial features more drastically in the similar direction to the outlier, which makes the detectors easier to detect. At the same time, a larger perturbation budget gives HFC more room to manipulate the deep representation and enhance its ability to compromise the adversarial detectors more drastically. 

On the other hand, as illustrated in Sec.~\ref{Sec:feature_vulnerable}, the Chest X-ray image classification models are more vulnerable than CIFAR-10 ones. As a consequence, HFC enjoys more success in manipulating the adversarial feature into the feature distribution of normal samples even with a small perturbation. Compared to CIFAR-10 natural images, when the attacker utilizes HFC to camouflage Chest X-ray medical adversarial attacks, greater performance drops of the adversarial detectors can be found in Table~\ref{Table:Cifar}. This discovery suggests that \textit{improving the robustness of medical models is necessary.}

\begin{figure}[t]
\begin{center}
   \includegraphics[width=1\linewidth]{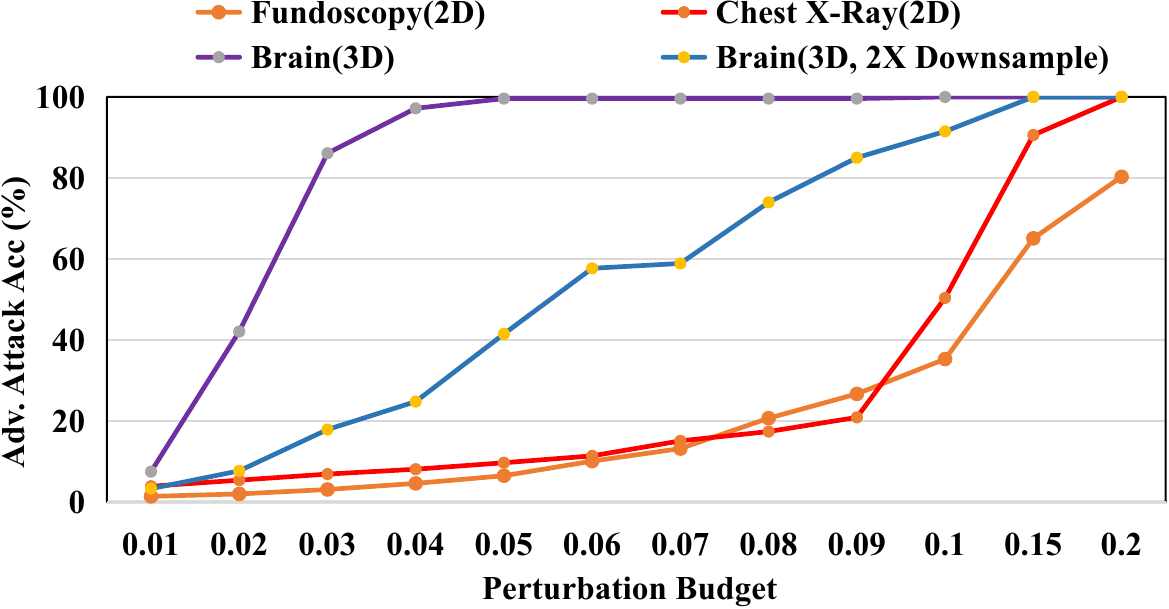}
\end{center}
   \caption{The comparison of the successful rate (Adv. Acc, $\uparrow$) of 3D and 2D medical images. We choose BIM to attack ResNet-18-3D, ResNet-50 for 3D, 2D images, respectively. Greater Adv. means more vulnerability.}
\label{Fig:3D_2D_advacc}
\vspace{-5mm}
\end{figure}

\subsection{Comparison 3D and 2D vulnerability}

In addition, we compare the vulnerability of 2D and 3D medical volumes. Specifically, we attack the disease classification network with BIM within tiny perturbation budgets ($L_{\inf} = [0.01\sim0.2]/256$). As shown in Fig.~\ref{Fig:3D_2D_advacc}, BIM has a greater successful rate on 3D volumes than 2D ones, especially within $L_{\inf} = 0.02/256$, which means the 3D medical volumes are likely more vulnerable than 2D images.

\bheading{Analysis:} According to Goodfellow's linear explanation~\cite{goodfellow2014explaining}, sufficient dimensionality is the primary source of CNN vulnerability to adversarial attacks. We speculate that increasing the dimensionality of input images likely makes neural networks more susceptible to adversarial perturbations, and the larger dimensionality of 3D images brings greater vulnerability. In our experiments, the input dimensionality for 3D images is $30\times270\times270$, while for 2D images, it is $256\times256\times3$, which is significantly smaller. We show that 3D input is more vulnerable. To better validate this hypothesis, we design an experiment to evaluate the robustness by reducing the dimensionality of the 3D input. As illustrated in Fig.~\ref{Fig:3D_2D_advacc}, when we 2x-downsample the 3D images and reduce the overall 3D input dimensionality by $8\times$, a significant improvement in adversarial robustness is observed, which somewhat confirms the linear explanation~\cite{goodfellow2014explaining} hypothesis.

\begin{table}[t]
\caption{The AUC scores (\%, $\uparrow$) represent the performances of HFC term to detect the OOD signals, which are various sizes and quantities of synthesized ``lesions''. FR (\%) denotes the false rates of the classification network influenced by OOD signals, leading to incorrect predictions.}
\centering
\small
\begin{tabular}{c|c|rr|rr}
\bottomrule \hline
\multicolumn{2}{c|}{Lesions}                        & \multicolumn{2}{c|}{Fundoscopy}                      & \multicolumn{2}{c}{Chest X-ray}                    \\ 
\hline
Size & Quantity & AUC(\%) & FR(\%) & AUC(\%) & FR(\%) \\
\hline
small &  2                   & 87.4                  & 0.2                   & 94.4                & 1.2                \\
small &  4                   & 92.4                  & 0.6                   & 97.7               & 4.3                \\
small &  6                   & 94.6                  & 0.7                   & 98.1               & 10.0                \\
\hline
medium &  1                   & 84.9                  & 0.2                  & 94.8                & 0.5               \\
medium &  2                   & 91.4                 & 0.2                  & 97.9                 & 1.2             \\
medium &  4                   & 95.5                 & 0.5                   & 99.5                 & 3.4               \\
\hline
large &  1                   & 91.9                  & 0.2                   & 98.1                & 2.1               \\
large &  2                   & 94.9                  & 0.4                  & 99.6              & 4.3                \\
large &  3                   & 97.2                & 0.5                   & 99.7                & 7.5               \\
\hline \toprule 
\end{tabular}
\label{Table:OOD}
\vspace{-5mm}
\end{table}

\subsection{Detecting OOD signals using HFC term}

In this section, we try to evaluate the possibility of using HFC term to detect the out-of-distribution (OOD) outlier data. Because obtaining the real-world OOD signals (\textit{e.g.}, artifacts, anatomical variations, or unseen pathologies) is difficult, we modify lesion synthesis methods proposed by Hu et al.~\cite{hu2023label,yao2021label} to simulate OOD images by synthesizing different sizes and quantities of ``lesions'' and insert into a random position in clean images, which are served as anomaly signals and shown in Fig.~\ref{Fig:ood}. Specifically, the ``lesion'' generation process employs random elastic deformations to deform randomly-sized ellipses and generate various shapes. Additionally, Gaussian filtering with random parameters is applied to filter salt noise to generate various textures, which are then incorporated into the ellipses. Finally, blurring is applied to the edges. 

Next, we respectively compute the anomaly scores $S_{\mathrm{HFC}}$ (in Eq.~\ref{Eq:anomaly}) for clean images and OOD images on the test set, as well as AUC scores accordingly. Given a certain lesion quantity and type, we randomly synthesize anomalies ten times for each image. For simplicity, we ignore the wrong predicted OOD images caused by the synthesized mask, the false-prediction rates (FR) caused by the OOD signals are shown in Table~\ref{Table:OOD}. Moreover, we change the sizes and quantities of lesions to control the degree of outlier. Surprisingly, as the results in Table~\ref{Table:OOD}, we find that our HFC does have the potential to detect the possible outliers with AUC scores larger than 85\%, even the small OOD signal such as small lesions. Furthermore, HFC exhibits a stronger detection capability for more potent OOD signals (larger and more numerous lesions). This finding inspires a strong OOD detection method for more real-world OOD signals in the future.

\subsection{Limitation}

It is a limitation of our method to directly apply on auto-PGD (APGD)~\cite{croce2020reliable}, which auto-decays the step size (also named as learning rate) of each iteration in PGD according to the gradient. Therefore, directly adding the HFC term on the original cross-entropy loss disrupts the original automatic decay mechanism of APGD.



    
    

\section{Conclusion}

In this paper, we investigate the characteristics of medical adversarial examples in feature space. It is theoretically proved that existing adversarial attacks tend to optimize the features in a similar direction. In consequence, the adversarial features become outliers and easy to detect. A stress test is then performed to reveal the greater vulnerability of medical image features, compared to the natural ones. However, this vulnerability can be exploited by the attacker to hide the adversarial features. We herein propose a novel hierarchical feature constraint (HFC), a simple-yet-effective add-on that can be applied to any existing attack, to avoid adversarial examples from being detected. The effectiveness of HFC is validated by extensive experiments, which also significantly surpasses other adaptive attacks. HFC reveals the limitation of the current defenses for detecting medical adversarial attacks in the feature space and allows to develop the importance of the robustness of the medical features. We hope it can inspire stronger defenses in the future.


\bibliographystyle{IEEEtran}
\bibliography{Main}


\end{document}